\documentclass[a4paper,twocolumn,11pt]{quantumarticle}
\pdfoutput=1
\usepackage[utf8]{inputenc}
\usepackage[english]{babel}
\usepackage[T1]{fontenc}
\usepackage{amsmath}
\usepackage{hyperref}
\usepackage{float}
\usepackage{caption}
\usepackage{tikz}
\usepackage{lipsum}

\usepackage{amsmath}
\usepackage{physics}
\usepackage{graphicx}
\usepackage{caption}
\usepackage{subcaption}
\graphicspath{ {./images/} }
\usepackage{xcolor}
\usepackage[normalem]{ulem}

\newcommand{\AR}{approximation ratio }

\DeclareMathOperator*{\argmin}{arg\,min}

\usepackage{xcolor}     
\usepackage[normalem]{ulem}

\newcommand{\betagamma}{(\vec{\beta},\vec{\gamma})} 
\newcommand{\betagammastar}{(\vec{\beta}^{*},\vec{\gamma}^{*})} 

\begin{document}

\title{Iterative-Free Quantum Approximate Optimization Algorithm Using Neural Networks}
\author{Ohad Amosy*}
\affiliation{Faculty of Computer Science, Bar-Ilan University, 52900 Ramat Gan, Israel}
\email{amosyoh@biu.ac.il}

\author{Tamuz Danzig*}
\affiliation{Faculty of Engineering, Bar-Ilan University, 52900 Ramat Gan, Israel} \email{tamuz.danzig@gmail.com}
\thanks{\newline* Equal contribution}

\author{Ely Porat}
\affiliation{Faculty of Computer Science, Bar-Ilan University, 52900 Ramat Gan, Israel} 

\author{Gal Chechik}
\affiliation{Gonda Brain research institute in Bar-Ilan University, 52900 Ramat Gan, Israel. } 
\affiliation{NVIDIA Research} 

\author{Adi Makmal}
\affiliation{Faculty of Engineering, Bar-Ilan University, 52900 Ramat Gan, Israel} 
\email{adi.makmal@biu.ac.il}

\maketitle

\begin{abstract}
The quantum approximate optimization algorithm (QAOA) is a leading iterative variational quantum algorithm for heuristically solving combinatorial optimization problems. 
A large portion of the computational effort in QAOA is spent by the optimization steps, which require many executions of the quantum circuit. 
Therefore, there is active research focusing on finding better initial circuit parameters, which would reduce the number of required iterations and hence the overall execution time. 
While existing methods for parameter initialization have shown great success, they often offer a single set of parameters for all problem instances. We propose a practical method that uses a simple, fully connected neural network that leverages previous executions of QAOA to find better initialization parameters tailored to a new given problem instance. 
We benchmark state-of-the-art initialization methods for solving the MaxCut problem of Erdős–Rényi graphs using QAOA and show that our method is consistently the fastest to converge while also yielding the best final result. 
Furthermore, the parameters predicted by the neural network are shown to match very well with the fully optimized parameters, to the extent that no iterative steps are required, thereby effectively realizing an   iterative-free QAOA scheme.
\end{abstract}

\section{Introduction}
Quantum computers are currently found in the so-called noisy intermediate-scale quantum (NISQ) era, comprising a small number of qubits and high error rates that accumulate rapidly with the number of operations \cite{preskill2018quantum}. 
A popular family of quantum algorithms in the NISQ era is the variational quantum algorithms (VQAs), which offer a heuristic approach for solving optimization problems \cite{cerezo2021variational}. VQAs use relatively shallow quantum circuits and are, therefore, more resilient to noise compared to standard quantum algorithms \cite{mcclean2017hybrid}. 
They employ quantum circuits of parameterized gates that are updated, by means of classical computation, to reach an optimum of a predefined objective function \cite{nannicini2019performance}. 
The quantum approximate optimization algorithm (QAOA) is a particular VQA designed specifically to solve approximate combinatorial optimization problems~\cite{farhi2014quantum, farhi2016quantum, crooks2018performance}, which are of high importance for various fields in both industry and academia~\cite{korte2011combinatorial, papadimitriou1998combinatorial}.

In VQAs and QAQA in particular, each optimization step involves thousands of circuit executions, so reducing  the number of optimization steps is highly desirable.
Choosing a proper QAOA parameters can improve both the convergence rate \cite{zhou2020quantum, sack2021quantum} and the accuracy of the final solution \cite{farhi2014quantum, sack2021quantum}.
Finding better initial parameters for the QAOA circuit is thus a matter of active research \cite{crooks2018performance, zhou2020quantum,sack2021quantum, brandao2018fixed, khairy2019reinforcement,    akshay2021parameter, alam2020accelerating,rabinovich2022progress}.

Most initialization methods offer a single, common, set of parameters for all problem instances \cite{crooks2018performance, brandao2018fixed, zhou2020quantum, akshay2021parameter}. 
This is a pragmatic approach and often very successful in reducing the number of required QAOA iterations. 
Nevertheless, it inherently ignores important relations between problem instances and their optimal parameters. These relations, which are based on the information carried by the particular description of each problem instance, can potentially be utilized to save further iterations.
Other methods that manage to personalize the initial parameters per problem instance \cite{sack2021quantum,alam2020accelerating} show an impressive improvement, exponentially better compared to a random parameter initialization \cite{sack2021quantum}, yet at a high cost of many additional quantum circuit executions. 
These findings entail that using a different initialization per problem instance is beneficial and motivate the exploration of more cost-effective schemes for generating personalized parameters for each problem instance. 

The goal of this paper is to reduce the number of required QAOA iterations 
without compromising  algorithm performance and without invoking any additional quantum circuit executions, by using better initialization parameters.

We propose a practical approach based on neural network (NN) that predicts appropriate initialization parameters for QAOA \textit{per problem instance}. The NN takes as input a predefined encoding of the problem instance and outputs the corresponding QAOA parameters. To that end, the NN is trained using past results of QAOA optimizations as labeled data. 
We evaluate our method on the standard MaxCut benchmark; The method can be similarly applied to other problems.

This study provides the following contributions:
(a) it demonstrates that the QAOA's variational parameters can be learned efficiently and robustly by a simple NN for the MaxCut problem without any additional quantum computation; 
(b) the proposed learning scheme is empirically shown to outperform state-of-the-art QAOA initialization methods in terms of the approximation ratio and convergence speed;  
(c) using the proposed method may relinquish the need for any further optimization of the quantum circuit altogether, potentially saving many costly quantum circuit executions; and 
(d) we show that these advantages become more profound as the size of the problem (the number of nodes in the graph) increases.

The paper is structured as follows: Sec.~\ref{sec:related works} provides the relevant background. It describes the QAOA algorithm, the MaxCut problem, and previous works on parameter initialization for the QAOA circuit. Sec.~\ref{sec:method} introduces our proposed method. 
Sec.~\ref{sec:results} benchmarks several initialization techniques alongside our method and provides a detailed comparison of their performances. Finally, we discuss the central advantages of the proposed method for the NISQ era in Sec.~\ref{Sec:Discussion}.

\section{Background}
\label{sec:related works}

\subsection{The quantum approximate optimization algorithm (QAOA)}\label{Sec:QAOA}
The QAOA can be regarded as a time-discretization of adiabatic quantum computation \cite{farhi2014quantum, crooks2018performance,farhi2000quantum}, whose $p$-layers circuit Ansatz constructs the following state:
\begin{equation}\label{eq:QAOA_Ansatz}
\ket{\psi_{p}\betagamma} = \left[\prod_{l=1} ^p e^{-i\beta_{l} H_M} e^{-i\gamma_{l} H_C}\right] \ket{+}^{\otimes N}
\end{equation}

\noindent
where $N$ is the number of qubits. $H_C$ is called the problem Hamiltonian, which is defined uniquely by the specific problem we are trying to solve (see below for the MaxCut problem example), and $H_M$ is called the mixer Hamiltonian, provided in the same form for all problems, and given by $H_M = \sum_{n=1}^{N} \sigma_{n}^{x}$, where $\sigma_{n}^{j}$ is the Pauli operator $j$ that acts on qubit $n$. Finally, $\ket{+}=\frac{1}{\sqrt{2}}(\ket{0}+\ket{1})$ is the $+1$ eigenstate of $\sigma^{x}$. The $p$-dimensional vectors $\vec{\beta}$ and $\vec{\gamma}$ are the variational real parameters that correspond to $H_M$ and $H_C$, respectively. The objective cost function is determined by the expectation value of the problem Hamiltonian
\begin{equation}\label{eq:QAOA_Cost}
F_{p}\betagamma = \bra{\psi_{p}\betagamma} H_C \ket{\psi_{p}\betagamma} \quad, 
\end{equation}
and we denote the optimal parameters by $\betagammastar$, with which the optimal solution is attained: 
\begin{equation}\label{eq:Best_Params}
\betagammastar = \textrm{arg} \;\underset{\vec{\beta},\vec{\gamma}}{\textrm{opt}}\;F_{p}\betagamma.
\end{equation}

QAOA is an iterative algorithm: it starts with an initial guess of the $\betagamma$ parameters, then the expectation value of the problem Hamiltonian of Eq.~\ref{eq:QAOA_Cost} is evaluated by repeated measurements of the same circuit to reach a certain level of statistical accuracy. Once the cost function is calculated, the $\betagamma$ parameters are updated towards the next iteration by a classical optimizer so as to optimize $F_{p}(\vec{\gamma},\vec{\beta})$. The overall QAOA iterative process requires many executions of the quantum circuit; first, reaching a satisfactory level of statistical error $\epsilon$ typically requires  $O(\epsilon^{-2})$ shots \cite{gilyen2019optimizing}. To keep the same level of accuracy, the number of shots grows with the system size; it  was recently estimated to grow exponentially with size \cite{qaoa_scaling_lotshaw2022}; 
second, the optimization process requires additional circuit executions: 
if the optimizer is gradient-based, 
then the derivative of the cost function with respect to the $\betagamma$ parameters must also be calculated at each iteration, requiring many additional executions of the quantum circuit to evaluate the gradients, e.g.,\ by following the so-called ``parameter shift rule" \cite{mitarai2018quantum, schuld2019evaluating}. Alternatively, gradient-free optimizers 
can avoid the calculations of the derivatives, yet at a high cost of many more optimization iterations \cite{conn2009introduction}. 
The iterative QAOA process thus requires many executions of the quantum circuit and each optimization step that can be avoided translates directly into a significant reduction in computational resources.

\subsection{The MaxCut problem}\label{Sec:MaxCut}
The maximum-cut (MaxCut) problem is an NP-hard combinatorial problem that has become the canonical problem to benchmark QAOA ~\cite{zhou2020quantum, willsch2020benchmarking}.
It is defined over an undirected graph $G=(V,E)$, where $V={1,2,...,N}$ denotes the set of nodes, and $E$ is the set of edges. The (unweighted) {\em maximum cut} objective is the partition of the nodes into two groups $\{+1,-1\}$, such that the number of edges connecting nodes from the two different groups is maximal. 

Within QAOA, the problem Hamiltonian $H_C$ that corresponds to the MaxCut problem is given by \cite{farhi2014quantum}: 
\begin{equation}\label{eq:MaxCut_Cost}
H_C = \frac{1}{2}\sum_{(i,j)\in E}  \left(1-\sigma_{i}^{z}\sigma_{j}^{z}\right) \quad,
\end{equation}
such that an edge $(i,j)$ contributes to the sum if and only if the $(i,j)$ qubits are measured anti-aligned. 
The common performance metric of QAOA for the MaxCut problem is the approximation ratio:
\begin{equation}\label{eq:AR}
r = \frac{F_{p}\betagammastar}{C_{max}}
\end{equation}
where $C_{max}$ is the maximum cut of the graph.

\subsection{QAOA initialization techniques}\label{sec:qaoa_initialization_techniques}
The attempt to find optimal parameters for QAOA, which would require a minimum number of optimization steps for solving the MaxCut problem, is currently an extensive research topic. 


The most intuitive initialization scheme is linear, where the $\betagamma$ parameters vary linearly from one layer to the other, such that $\beta$ is gradually turned off and $\gamma$ is turned on. This linear approach is inspired by adiabatic quantum computation~\cite{farhi2000quantum}: the circuit begins from the ground state of the mixer Hamiltonian and aims at reaching the ground state of the problem Hamiltonian. While very simple and computationally efficient, the linear solution can be sub-optimal, see e.g.,\ \cite{zhou2020quantum, willsch2020benchmarking}.  As an alternative approach, several initialization methods selected the same set of parameters for common type graph instances, such as regular graphs \cite{crooks2018performance, brandao2018fixed,zhou2020quantum, akshay2021parameter,shaydulin2019multistart}. For example, it was suggested in \cite{crooks2018performance} to use a batches optimization method, in which initial parameters are found by optimizing batches of graphs in parallel. This way, the parameters fit multiple graphs and should also fit new graphs. The problem, however, with such homogeneous methods is that in practice, the optimal parameters change from one graph instance to another, even within the same family of graphs \cite{zhou2020quantum}. Such non-personalized methods ignore by construction possible relations between specific problem instances and their corresponding optimal parameters. 

Indeed, other studies have suggested using a different, personalized set of parameters for each problem instance. One method suggested initializing a $p+1$ layers circuit based on the optimized $p$ layers circuit. This proved useful \cite{zhou2020quantum}: for 3-regular graphs, the number of optimization steps reduced exponentially, from $2^{O(p)}$ for random parameter initialization to $O(poly(p))$ \cite{zhou2020quantum}. However, it required the full optimization of the $p$ layers circuit. Similarly, Ref.\ \cite{alam2020accelerating} used a regression model to predict the initial parameters of QAOA for a circuit with $p \ge 2$, given the optimized parameters for the corresponding single layer circuit, i.e.\ $p=1$. This requires the generation of an optimized dataset for both the one-layer circuit and the desired $p$-layers circuit; Moreover, given a new problem instance,  this method  requires full optimization of its single-layer circuit. 
Recently, it was shown that reusing the optimal parameters of small graph instances can be beneficial for solving the MaxCut problem of larger graph instances, when using single-layer QAOA circuits ($p=1$) \cite{galda2021transferability}. This is due to the  confined combinatorial number of small regular subgraphs with radius $p$ (the connection between the number of circuit's  layers and the performance on $p$-radius graphs was analyzed in \cite{farhi2014quantum} and is also addressed here, in Sec.\ \ref{sec:graph-size}). However, for $p>1$ the combinatorial number of subgraphs increases rapidly and the method becomes less practical.


The Trotterized quantum annealing (TQA) is an initialization protocol that personalizes the initial parameters for a new problem instance by running the quantum circuit on many different linear initializations and choosing the initial parameters that achieve the best result \cite{sack2021quantum}. Another approach proposed in \cite{shaydulin2019multistart} is a multi-start method in which the objective function is evaluated in different places in the parameters domain by using gradient-free optimizers. They show that it is possible to reach good minima with finite number of circuit evaluations. 
These methods perform well but at a high computational cost per problem instance, as they optimize the initial parameters for each new problem instance. In particular, they do not leverage the knowledge obtained from previous QAOA optimizations.


Several methods used machine learning techniques to improve the performance of QAOA. 
For example, several studies have attempted to facilitate the optimization step by replacing the standard classical optimizers  with reinforcement learning methods and meta-learning algorithms~\cite{khairy2019reinforcement,wauters2020reinforcement, khairy2020learning, khairy2019reinforcement2,  wilson2021optimizing, verdon2019learning, wang2021quantum}. 
Egger et al.\ proposed the warm-start QAOA algorithm, where the qubits are initialized in an approximated solution instead of the uniform superposition as typically done in QAOA~\cite{egger2021warm}.
Jain et al.\ used a graph neural network (GNN) for predicting the probability of each node being on either side of the cut in order to initialize warm-start QAOA~\cite{jain2021graph}. 
Unlike these methods, we use NN only to predict the initial parameters. Since we do not alter the optimization process, our method can be applied together with any of the other methods.

\section{The Method}\label{sec:method}
\begin{figure}[h!]
\centering
\includegraphics[width=3.3in]{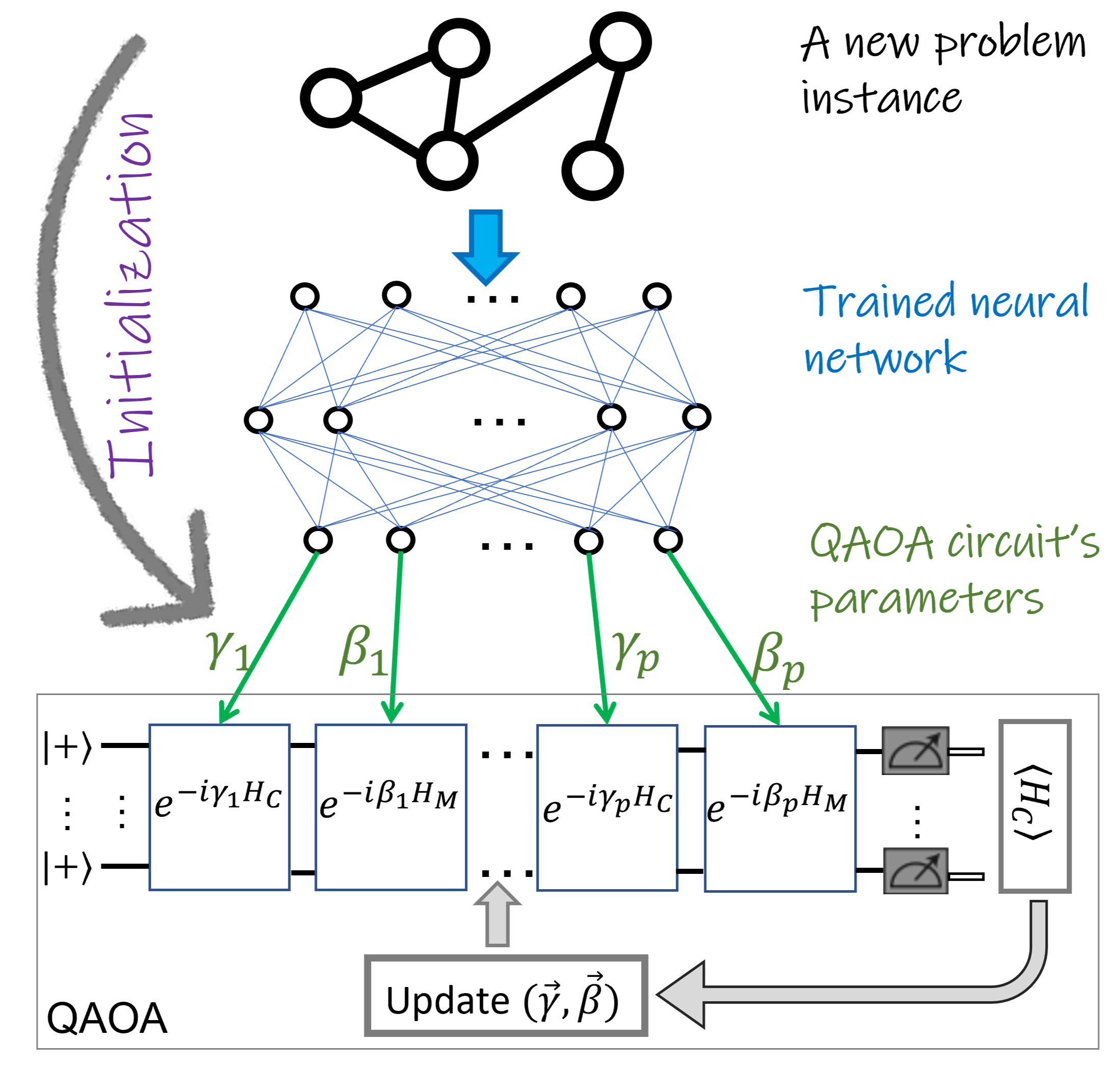}
\caption{An Illustration of the proposed method: 
each new problem instance (e.g., a graph in the MaxCut problem) is encoded (e.g., by its adjacency matrix) and inserted as an input into a trained neural network. The latter predicts better initial parameters for QAOA, specifically tailored to the given problem instance (the graph).}
\label{fig:method_illustration}
\end{figure}

Given a combinatorial optimization problem, we predict the best initialization parameters for QAOA for a new particular problem instance, as illustrated in Fig.~\ref{fig:method_illustration}. 
We use a simple, fully connected neural network (NN) that takes as an input an encoding of the problem instance and predicts an initialization parameters for that specific instance. The encoding of the problem instances may vary from one combinatorial problem to another. In our case of solving the MaxCut problem we encode each graph instance by its adjacency matrix, which is then fed to the network, as described below.

The method is based on the assumption that the QAOA was already applied on $n$ different problem instances $S=\{s_i\}_{i=1}^n$, and that both the set of instances and the set of the corresponding final parameters $P=\{\vec{\beta_i},\vec{\gamma_i}\}_{i=1}^n$ were saved and can be used as labeled data for the NN training. This allows us to exploit past QAOA calculations for predicting the initial parameters $(\vec{\beta}_{new},\vec{\gamma}_{new})$ for a new problem instance. 

Formally, we train a neural network $f_\theta$ to map each problem instance from $S$ to its optimal circuit parameters from $P$. The input of the NN encodes the problem instance\, and the NN output is a vector of size $2p$, where $p$ is the number of layers in the quantum circuit: $p$ parameters for $\vec{\beta}$ and $p$ parameters for $\vec{\gamma}$. The neural network parameters $\theta$ are trained to minimize the $L2$ norm
\begin{equation}\label{eq:nn_loss}
\theta = \argmin_\theta \sum_{i=1}^n ||f_\theta(s_i) - (\vec{\beta_i},\vec{\gamma_i})||_2.
\end{equation}
Finally, once the neural network is trained, predicting the initial parameters for a QAOA calculation of a new problem instance is done by setting $(\vec{\beta}_{new},\vec{\gamma}_{new}) = f_\theta(s_{new})$.

In comparison to previous methods, which either offer a fixed set of parameters for all problem instances or personalize the parameters per problem instance, but at a high cost of extra quantum circuit execution, our scheme manages, by design, to personalize the parameters per problem instance with no such extra cost. This is an \textit{a-priori} advantage. In the next section the empirical performance of the methods is examined.

\subsection{Applying to the MaxCut problem}
We exemplify our method on the MaxCut problem. We encode each problem instance, i.e., a graph, by its adjacency matrix. The matrix is reshaped into a one-dimensional vector of length $\frac{N(N-1)}{2}$, which is given as an input to the NN. 

\section{Results}\label{sec:results}
\subsection{Setup, benchmarking, and implementation details}\label{sec:Benchmarking}
\textbf{Setup} \\
We test the proposed method by solving the MaxCut problem for Erdős–Rényi graphs of $N=6 - 16$ nodes. In an Erdős–Rényi graph, each edge $(i,j)$ exists with a certain probability, independently from all other edges. We consider two different graph ensembles: (a) \textbf{constant Erdős–Rényi graphs}  - where the existing probability for each edge is constant and equals $p=0.5$, for all graphs; (b) \textbf{random Erdős–Rényi graphs} - where the edge probability changes between graphs but remains the same  for all edges within a graph: for each graph, we uniformly sample a probability in the range $p\in\{0.3,0.9\}$ and assign it to all the edges. 
While the first, constant probability setup is the most commonly considered in the literature, see e.g., \cite{sack2021quantum}, the latter setup with a random edge probability, better represents natural, real-world scenarios, where the probability for creating a connection between nodes depends on variables that may vary across graphs. For example, if the nodes of the graph represent people and a connection indicates an interaction between two people, then the probability of the interaction usually varies depending on external variables, such as place, age, culture, etc., and varies between different communities.
\\
\\
\textbf{Benchmarking} \\ 
To evaluate the proposed initialization method, we compare its performance to those of four initialization methods: (a) the batches optimization method \cite{crooks2018performance}; (b) the Trotterized quantum annealing (TQA) initialization procedure~\cite{sack2021quantum} with a predefined $\Delta t$; (c) a simple linear method; and (d) an average method. 
The first two methods \cite{crooks2018performance,sack2021quantum} are state-of-the-art methods, described in  Sec.~\ref{sec:qaoa_initialization_techniques}, whereas the latter two are simpler yet natural baselines. We chose to focus on the batches and the TQA optimization methods because they are the best initialization methods known to date that, given a new test graph, do not require any additional computational effort, as is also the case in our proposal. 

To enable the comparison, we implemented all four benchmark methods, as follows.
In the batches optimization method, the QAOA initial parameters are determined by finding a fixed, optimized set of parameters for a large training set of graphs that are optimized in parallel. This set of optimized parameters is then used as the initialization point for new test graphs. We implemented the batches optimization method, where we used 200 training graphs for each setup choice, i.e., \{graph ensemble (random or constant ER), number of qubits, and number of layers\}. 
This is the same training set size used in Ref.~\cite{crooks2018performance}, which we explicitly verified is sufficient, in the sense that optimizing over a larger set did not yield parameters that performed better on the test set.

The TQA initialization procedure takes the following linear form: 
\begin{equation}\label{eq:TQA}
\beta_l = \left(1-\frac{l}{p}\right)\Delta t, \quad   \gamma_l = \frac{l}{p}\Delta t,
\end{equation}
where $l=1...p$ 
indicates the layer's number in the quantum circuit. In the complete TQA protocol, the hyperparameter $\Delta t$ is determined by performing a simple grid search that optimizes the overall performance for each graph. This requires extra quantum circuit executions for each new graph instance, which we wanted to avoid to enable a fair comparison. It was numerically observed in Ref.~\cite{sack2021quantum} that for graph ensembles of similar nature (e.g.,\ regular unweighted graphs, weighted regular graphs, and constant ER graphs) the optimal value of $\Delta t$ does not vary by much. Therefore, in our experiments, for each kind of ensemble (constant and random ER), number of nodes $n$, and number of layers $p$, we performed the complete TQA protocol over 50 training graphs and averaged their results to find a single $\Delta t^*$ that optimizes the overall performance. We then used this optimized $\Delta t^*$ to evaluate the performance of TQA on new graphs. See Sec.~\ref{sec:app-tqadt} in the appendix for more details. 

We also implemented the simpler baselines: (a) a linear initialization in which $\beta$ decreases linearly from $\frac{\pi}{4}(1-\frac{1}{p})$ to $\frac{\pi}{4p}$ and $\gamma$ increases linearly from $\frac{\pi}{p}$ to $\pi(1-\frac{1}{p})$. 
This schedule is similar to the TQA one but simpler, as it employs no further hyper-parameters. The reason for this particular choice is the periodicity of $\beta_i \in [0,\frac{\pi}{2}]$ and $\gamma_i \in [0,2\pi]$ when plugged into Eq.\ \ref{eq:QAOA_Cost}. 
Note that a similar choice was made in \cite{zhou2020quantum, fuchs2021efficient}; (b) an average initialization, which takes the optimal parameters of 100 training graphs and averages them. 
This initialization method has not yet been addressed in the literature. Yet, it is natural to consider it once we have independent sets of optimized $\betagamma$ parameters, as we do in this work. Moreover, our numerical results indicate that, given its simplicity and performance, this is a relatively good choice.  
We further tried a random initialization scheme, but it performed poorly, starting at a low approximation ratio and reaching suboptimal local minima after optimization. Thus, it is not shown here.  
\\
\\
\textbf{Implementation Details} \\
We built a dataset of $5,000$ graphs for each setup combination, namely the size of the graph, number of layers, and sampling technique, as described above. 
In particular, we considered Erdős–Rényi (ER) graphs of sizes 6 to 16 nodes, whose edges are sampled randomly at a probability of 0.5 or using a random uniform probability in the range $[0.3, 0.9]$. Then, we optimized each graph's MaxCut solution by the QAOA algorithm, using the BFGS classical optimizer \cite{broyden1970convergence, fletcher1970new, goldfarb1970family, shanno1970conditioning}, where we used the full TQA algorithm to initialize our $\betagamma$ parameters. 
Using this procedure, we built a labeled training set of \{graph, optimized $\betagamma$ parameters\} pairs, with which we trained the neural network. All our QAOA circuit executions were performed on Qiskit's noiseless statevector simulator \cite{aleksandrowicz2019qiskit}.

\begin{figure*}
     \begin{subfigure}{0.5\textwidth}
         \centering
         \includegraphics[width=\textwidth]{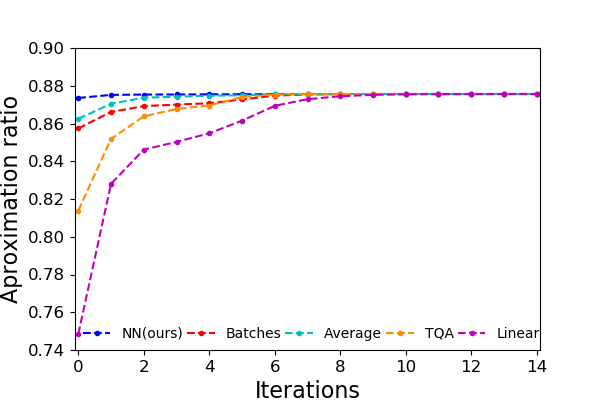}
         \caption{Constant ER with edge probability $p = 0.5$}
         \label{fig:converge_const}
     \end{subfigure}
     \begin{subfigure}{0.5\textwidth}
         \centering
         \includegraphics[width=\textwidth]{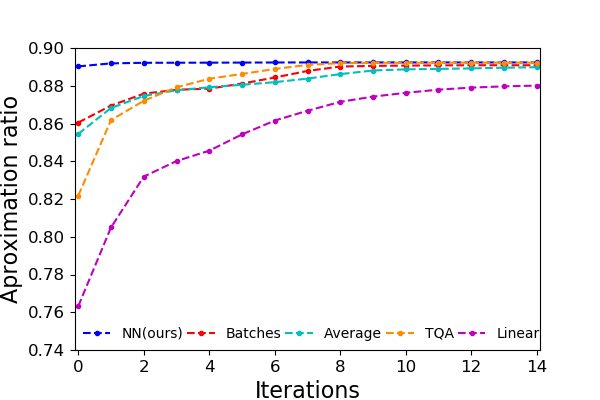}
         \caption{Random ER with edge probability $p \in [0.3,0.9]$ }
         \label{fig:converge_random}
     
     \end{subfigure}
     \caption{
     The approximation ratio during QAOA optimization, starting from initial parameters obtained by the different methods, averaged over 50 graphs of $N=14$ nodes.
     The QAOA circuit is composed of two layers.
     All graphs are sampled using ER: (a) with constant edge probability; (b) with random edge probability. The proposed NN method converges the fastest in both cases. A more significant improvement is observed in the case of graphs with random edge probability.}
    \label{fig:converge_const_random}
\end{figure*}

For all setups, we used the same simple 3-layer network architecture: the input layer is composed of $\frac{N(N-1)}{2}$ neurons, so as to encode the adjacency matrix of an undirected graph of $N$ nodes, the hidden layer has 100 neurons, and the output layer has $2p$ neurons that encode the learned $\betagamma$ parameters of the optimized QAOA circuit.

\subsection{Constant Erdős–Rényi}\label{sec:convergence}
We begin with ER graphs with $N=14$ nodes generated with a constant edge probability of 0.5. 
Fig.~\ref{fig:converge_const} compares the performance of the proposed NN method with those of: 
(a) the batches optimization method \cite{crooks2018performance}; (b) the TQA initialization procedure~\cite{sack2021quantum}; (c) the linear method, and (d) the average method,  
as described above. Each point in the graph indicates the averaged approximation ratio achieved by a 2-layer QAOA circuit over 50 test graphs 
as a function of the QAOA optimization step iterations. The standard error of the mean is less than $0.47\%$ and is not displayed in the figure for visual clarity.

It is seen that prior to the QAOA optimization steps, at the 0'th iteration, each method reaches a different approximation ratio and that all methods improve from one iteration to another until converging to the same approximation ratio. 
Yet, each method converges at a different rate.   
Fig.~\ref{fig:converge_const} demonstrates that, compared to all other methods, the NN method begins at a better \AR and converges the fastest. 
This means that the NN learns to predict adequate  $\betagamma$ parameters, given the adjacency matrix of a test graph.  
In fact, it can be observed that the initial \AR performance of our NN method is so close to the final parameters that hardly any optimization step is required.

It is also observed that the simple average method is the second best approach for the constant ER ensemble with an edge probability of $p=0.5$. This implies that the distribution over the $\betagamma$ parameters is rather confined. 
What would be the effect of a more widespread  ensemble distribution on our results? 
To check that, we next examine the performance of the benchmark initialization methods, alongside the NN one, on a random ER ensemble.

\subsection{Random Erdős–Rényi}\label{sec:random}
So far we tested our method in the standard setup: graphs are randomly drawn from ER with an edge probability of 0.5. We now expand this setup 
by assigning each graph with an edge probability that is sampled uniformly from the interval [0.3,0.9].
This way, we increase the diversity between the graphs, making it more realistic and challenging to guess good initial parameters \cite{zhou2020quantum, sack2021quantum, galda2021transferability}. 

Fig.~\ref{fig:converge_random} shows the performance of all our benchmark methods in the case of such a random-ER graph ensemble. The results resemble those of the constant-ER ensemble, with similar trends. 
In particular, the NN method begins from the highest \AR and converges faster than all other  methods. 
Yet, in the random-ER case, the differences between the methods are much more pronounced. 
First, it is seen that, in contrast to the constant-ER case, not all methods converge to the same approximation ratio, with the simple linear method reaching the lowest approximation ratio. Moreover, all methods, except for the NN one, exhibit a much slower convergence rate than the constant-ER scenario. 
The performance of the NN method is manifested in the achievable \AR at the zero'th iteration, showing a significant \AR gap of roughly $3\%$ from the second-best result obtained by the batches optimization method. 

This is not surprising: the NN method is designed to predict the best $\betagamma$ parameters per graph instance, i.e.,\ to personalize its performance. This is in contrast to all the benchmark methods, which essentially suggest a fixed set of $\betagamma$ parameters for all the test graphs, independent of the particular graph's structure. It should be noted that, in principle, the TQA has the capacity of personalizing over a particular graph, by searching for the optimal $dt$ that maximizes its results. However, as explained in Sec.~\ref{sec:Benchmarking}, this search of $\Delta t$ for each graph comes at a very high computational cost per graph, which we aim to avoid.

It is seen in Fig.~\ref{fig:converge_random} that also in the more challenging case of random-ER graphs, the NN method requires merely a single iteration to converge. In contrast, the TQA, which converges the fastest out of all other methods, requires about 8-12 iterations (depending on accuracy) to reach the same level of approximation ratio.  
We demonstrate the computational saving of our method by estimating the quantum circuit executions in our experiments. 
In our case, using the BFGS optimizer, we observed that even in the small $p=2$ circuit depth, each intermediate iteration requires, on average, 12 estimations of the cost function $F_p\betagamma$, each with different $\betagamma$ parameters, so as to evaluate the relevant gradients. As each cost function estimation typically requires thousands of shots to get a meaningful statistical accuracy, we get an overall saving of hundreds of thousands of quantum circuit executions. This saving is non-negligible and is expected to grow linearly with the circuit's depth.

\subsection{Varying the size of the graph}\label{sec:graph-size} 
\begin{figure}[h!]
\centering
\includegraphics[width=0.5\textwidth]{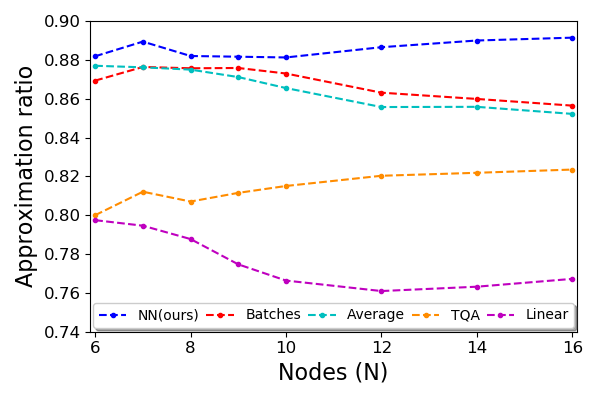}
\caption{The approximation ratio attained by each initialization method shown as a function of the number of nodes in the graph. Graphs were sampled using ER with a random edge probability for each graph size as in Fig.~\ref{fig:converge_random}. The approximation ratios were calculated at the zeroth iteration, i.e.,\ no optimization steps were taken. The number of layers is $p=2$. It is observed that the relative performance of our proposed method increases with graph size.}
\label{fig:graph_size}
\end{figure}

In previous sections we examined the performance of the proposed NN method on graphs with a fixed number of $N=14$ nodes. Next, we examine the method on random-ER graphs with varying number of nodes and draw our attention to the first QAOA evaluation, without applying any optimization steps.
To that end, we generated 200 different graphs for each graph size, from $N=6$ to $16$. Then, for each graph, we obtained the initial parameters using all the different methods, and ran a 2-layer QAOA circuit using those parameters (zeroth iteration only). 
Fig.~\ref{fig:graph_size} shows the average \AR achieved by each method as a function of graph size.

It is seen that the NN-method outperforms all other methods, irrespective of the graphs' size. Moreover, the preference of the NN-method over the other methods increases with the number of nodes, especially with respect to the batches and the average optimization methods. In comparison to the TQA method, the NN shows a rather fixed, large performance gap of roughly $8\%$ in \AR in favor of the NN method.

These results indicate that as we increase the graph size $N$ for a fixed number of layers $p$, the advantage of our method becomes more significant. 
To understand this trend, we go back to the structure of the QAOA algorithm and consider what happens when the QAOA circuit is kept fixed while the number of nodes increases, in approximating the MaxCut problem. We recall that QAOA is a $p$-local algorithm: in the case of the MaxCut problem, the algorithm's objective function, see Eq.~\ref{eq:QAOA_Cost}, is a sum of the expectation value terms of all edges in the graph (i.e., sum of $\bra{\psi_{p}(\vec{\gamma},\vec{\beta})}\sigma_{i}^{z}\sigma_{j}^{z}\ket{\psi_{p}(\vec{\gamma},\vec{\beta})}$ terms); following the commutation relation of the involved operators reveals that each expectation value term is influenced only by nodes that are at most $p$ edges away from the calculated edge term, i.e.\ all subgraphs with radius of at most $p$  \cite{farhi2014quantum,barak2021classical}. 
Zhou et al.\ showed that for certain families of graphs, such as the 3-regular graphs, where the possible number of subgraphs with confined radius $p$ is finite, the spread of the optimal parameters vanishes in the limit $N\rightarrow \infty$ \cite{zhou2020quantum}. This may justify a non-personalized approach which initializes a single set of $\betagamma$ parameters for all graph instances. In contrast, in random ER graphs, the number of confined subgraphs is not finite, but rather grows with $N$. Accordingly, the set of optimal parameters is not expected to converge to a single set as $N$ grows. This explains the decline of the approximation ratio as $N$ grows for methods that try to initialize with the same parameters for all graphs and emphasises the power of a personalized approach as proposed here. 
Next, we examine further the effect of personalizing the initial parameters.


\subsection{Personalization}\label{Sec:Personalization}

\begin{figure*}
     \begin{subfigure}{0.5\textwidth}
         \centering
         \includegraphics[width=\textwidth]{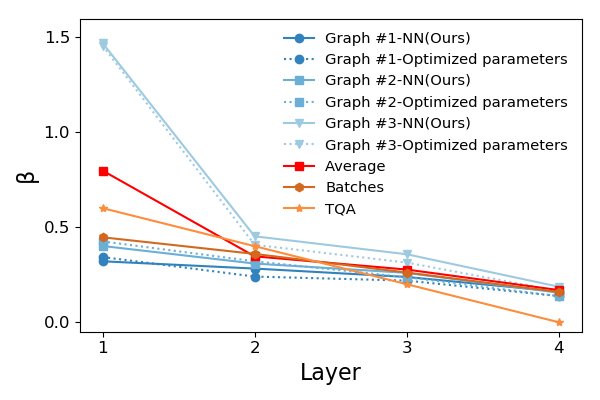}
         \label{fig:beta}
     \end{subfigure}
     \begin{subfigure}{0.5\textwidth}
         \centering
         \includegraphics[width=\textwidth]{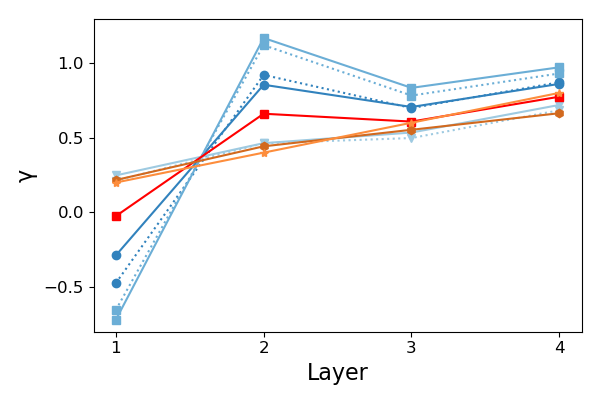}
         \label{fig:gamma}
     
     \end{subfigure}
     \caption{
     Initial $\beta$ (left figure) and $\gamma$ (right figure) parameters as a function of layers in the QAOA circuit, for three different individual graphs. A comparison is made between the prediction of the NN (solid blue lines), the optimized parameters (dashed blue lines), and the other non-personalized baselines: the simple average method (solid red line with square marks), the Batches method (solid brown line with circle marks), and the TQA (solid orange with star marks). The NN predictions are shown to follow closely the optimized curves for all three graphs.}
        \label{fig:personalization}
\end{figure*}

The key objective of our method is to create personalized initial parameters per problem instance. 
Fig.\ \ref{fig:personalization} depicts the $\betagamma$ values as a function of the layers in a 4-layers circuit for three individual graphs (graph-$1$, graph-$2$, graph-$3$) with $N=12$ nodes, sampled from the random ER ensemble.
The exact optimized parameters are marked in dotted-blue lines of different shades and different markers: graph-1 is marked in dark blue circles, graph-2 in blue squares, and graph-3 is shown in light blue triangles. The NN parameters are shown in solid lines with corresponding shades and markers. 
The simple average method, the batches-method, and the TQA, all produce a single set of parameters, depicted in red, brown, and orange, respectively.    
It is evident that the $\betagamma$ parameters obtained by our NN approach closely resemble the optimal parameters \textbf{per graph}, thus performing the personalization successfully. In contrast, other approaches, that do not have the flexibility to personalize the initial parameters, predict parameters that are approximately linear and 
are further away from the optimal parameters of the individual graphs.

     

\section{Discussion and Outlook}\label{Sec:Discussion}
This work joins an existing effort to find proper initialization for the QAOA  circuit's variational parameters. We showed that a simple neural network could be trained to predict initial $\betagamma$ parameters for QAOA circuits that solve the MaxCut problem, per graph instance. Moreover, we showed that these predicted parameters match very well with the optimal parameters, to which the QAOA iterative scheme eventually converges. This enables an effectively iterative-free approach, where the QAOA circuit is executed only once, with the predicted parameters, thus saving significant amount of computational resources. 
We demonstrated that our approach requires up to 85\% fewer iterations compared to current state-of-the-art initialization methods to reach optimized results for solving the MaxCut problem on both constant and random Erdős–Rényi graphs. 

Our method assumes the availability of previous QAOA optimizations for different instances of the same problem. The neural network we employed is a simple, fully-connected 3-layer network and the classical burden of training the network is negligible. Given a new problem instance, our method directly predicts the corresponding variational parameters without needing to execute the quantum circuit. In addition, as the training set grows, the neural network can rapidly adjust to the new data by few classical-learning iterations. Thus, in contrast to other methods, our method does not require any auxiliary executions on quantum devices. 
In this paper we employed the simplest deep network possible. Other NN architectures, like graph neural networks (GNN), may be better for this problem, and we expect this to be a topic of future research.

The ability of the proposed NN method to personalize the predicted parameters per graph instance  becomes more significant as the graph ensemble is more varied and the spread of optimal parameters is broader. We demonstrated this property by showing that while the proposed method outperforms all benchmark methods for constant-edge Erdős–Rényi graphs, it suppresses them even further when the edge probability is taken to be random. 
We thus conjecture that our method will provide an even more significant benefit for classes of graphs that show larger distributions, e.g., for random and weighted graphs, for which the optimal parameters are known to have a wider distribution \cite{zhou2020quantum}. Such graphs are prevalent in many practical applications, such as VLSI degisn, and social networking, see e.g., \cite{spirakis2021max}. 

Another manifestation of the same observation is that for the realistic scenario of a finite number of layers $p$ and a growing number of graph nodes $N$, our method leads to better performance compared to other methods (see Fig.~\ref{fig:graph_size}) for approximating the MaxCut of random Erdős–Rényi graphs; 
As quantum computers develop and have more qubits, they are able to solve larger instance problems. Yet, for practical solutions, the number of layers in QAOA must be finite. This makes our method especially practical in the NISQ era, where quantum devices increase in size but are too noisy for executing deep circuits.

Finally, in this work we focused on the MaxCut problem but we believe that our method is beneficial for any optimization problem, solved via a variational quantum algorithm (not necessarily QAOA), that incorporates an underlying mapping between the different instances of the problem and their corresponding optimized variational parameters. 

\section*{Acknowledgement}
We would like to thank Yoni Zimmermann and Yehuda Naveh for the fruitful discussions.

\bibliography{QuantumFirstLibrary}
\bibliographystyle{unsrt}

\clearpage
\onecolumn
\appendix

\section{Finding an optimal $\Delta t$ for the TQA method} \label{sec:app-tqadt}
\begin{figure}[h]
\centering
\includegraphics[width=3.6in]{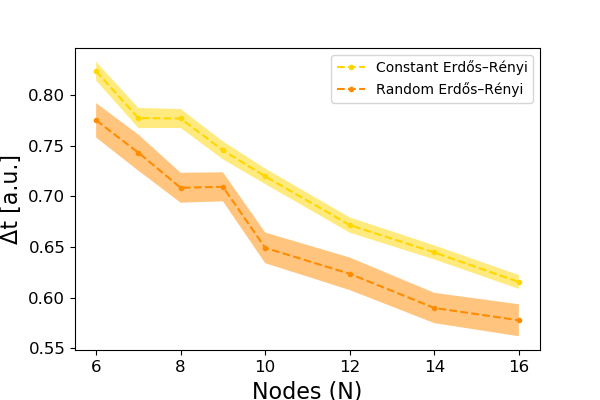}
\caption{Optimal $\Delta t$ for TQA initialization for each graph size (i.e.\ the number of nodes), with $p=2$. Constant Erdős–Rényi graph ensemble is drawn in yellow and random Erdős–Rényi graph ensemble in orange. The shades mark the standard error of the mean (SEM).}
\label{fig:tqadt_with_sem}
\end{figure}

In our experiments, for each number of nodes $N$, number of layers $p$, and graph ensemble (constant and random ER graphs, as in the main text), we applied a grid search to find the optimal $\Delta t$ (Eq.\ \ref{eq:TQA}), as in Ref.\ \cite{sack2021quantum}. We averaged the best $\Delta t$'s over 50 different graphs to obtain a single initialization that would fit as well as possible the variety of graphs. 
Fig.~\ref{fig:tqadt_with_sem} shows the optimal $\Delta t$ for different graph sizes, with $p=2$. For the random ER graphs, the spread of the optimal $\Delta t$ is wider, as the distribution of the graphs is broader. 

\end{document}